\documentclass[slac_one]{revtex4}
\usepackage{graphicx}
\usepackage{fancyhdr}
\newcommand{\kpi}{$K\pi^-$}
\newcommand{\mkpi}{$m_{K\pi^-}$}
\newcommand{\psipi}{$\psi\pi^-$}
\newcommand{\jpsipi}{$J/\psi\pi^-$}
\newcommand{\psitwospi}{$\psi(2S)\pi^-$}

\newcommand{\Ksone}{$K^{\ast}(892)$}
\newcommand{\Kstwo}{$K^{\ast}_2(1430)$}

\newcommand{\z}{$Z(4430)^-$}
\newcommand{\costhk}{$\cos\theta_K$}

\input{babarsym}

\begin{document}

\title{Searches for exotic $X$,$Y$, and $Z^-$ states with \babar\/}

\author{Arafat Gabareen Mokhtar (for the \babar\ Collaboration)}
\affiliation{SLAC, Stanford, CA 94025, USA}
\begin{abstract}
Recently, several charmonium-like states above $D\bar{D}$ threshold
have been discovered at the BELLE and \babar\ $B$-factories. Some of
these states are produced via Initial State Radiation ({\it e.g.}
$Y(4260)$ and $Y(4350)$) and some are observed in $B$ meson decays
({\it{e.g.}}  $X(3872)$, $Y(3940)$). The BELLE observation of the
enhancement in the $\psi(2S)\pi^-$, {\it {i.e.}} the \z\ state, has
generated a great deal of interest, because such a state must have
minimum quark content ($c\bar{c}d\bar{u}$), so that it would represent
the unequivocal manifestation of a four-quark meson state. Here we
report recent \babar\ results on the $Y(4260)$, $X(3872)$, $Y(3940)$,
and a search for the \z\/.
\end{abstract}

\maketitle

\thispagestyle{fancy}

\section{THE \mbox{\boldmath{$Y(4260)\rightarrow J/\psi\pi^+\pi^-$}}} 
\begin{figure*}[!htbp]
  \begin{center}
    \setlength{\unitlength}{1.0cm}
    \begin{picture}(10,9.5)
      \put(-2.,0){\includegraphics[height=0.4\textheight,width=0.7\textwidth]{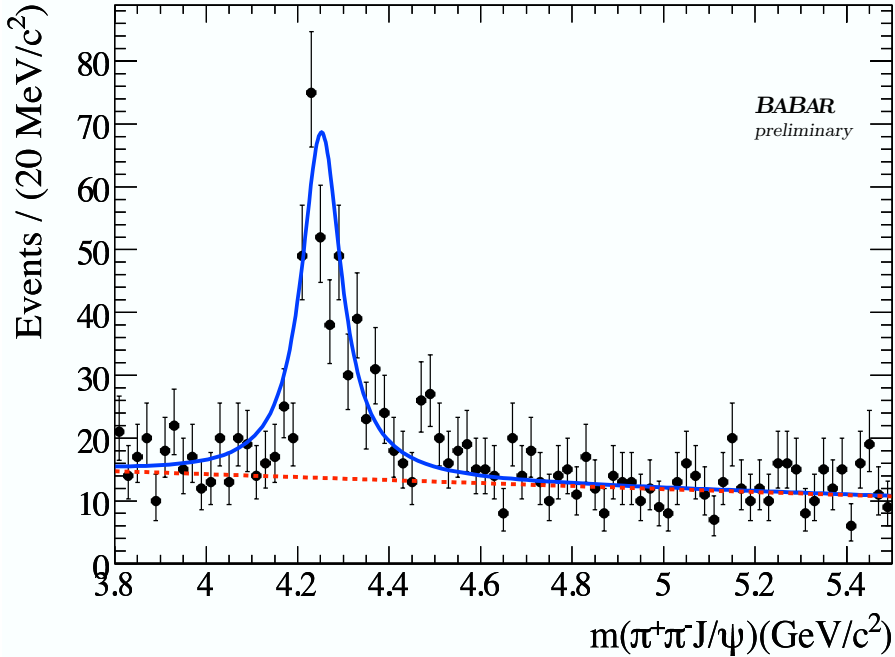}}
      \put(8.,7.5){\bfseries \babar}
      \put(8.,7.2){\scriptsize \bfseries {\sl preliminary}} 
    \end{picture}
    \caption{The ISR-produced $J/\psi\pi^+\pi^-$ invariant mass
      distribution. The dots represent the data, the solid curve shows the
      fit result, while the dashed curve represents the background
      contribution.}
  \end{center}
  \label{Y4260}
\end{figure*}

The $Y(4260)$ was discovered~\cite{Aubert:2005rm} by \babar\ in the
Initial State Radiation (ISR) process $e^+e^-\rightarrow
\gamma_{ISR}Y(4260)$, $Y(4260)\rightarrow J/\psi\pi^+\pi^-$. Being
formed directly in $e^+e^-$ annihilation, this should be a
$J^{PC}=1^{--}$ state. However, its nature is still not understood,
and does not seem compatible with a charmonium interpretation. In
Fig.~\ref{Y4260}, we show the ISR-produced $J/\psi\pi^+\pi^-$
invariant mass distribution for the full \babar\ dataset (454
fb$^{-1}$)~\cite{:2008ic}, where a clear enhancement of the $Y(4260)$
is observed. The updated $Y(4260)$ mass and width values are
$m=4252\pm 6(stat)^{+2}_{-3}(syst)$ \mevcc\/ and $\Gamma=105\pm
18(stat)^{+4}_{-6}(syst)$ \mev\/, respectively. There is no evidence
for the enhancement at $\sim 4005$ \mevcc\/ reported by the BELLE
Collaboration~\cite{:2007sj}.

\section{THE \mbox{\boldmath{$X(3872)\rightarrow J/\psi\gamma$}} AND  \mbox{\boldmath{$X(3872)\rightarrow \psi(2S)\gamma$}}} 
\begin{figure*}[!htbp]
  \begin{center}
    \setlength{\unitlength}{1.0cm}
    \begin{picture}(13,7.5)
      \put(-2.,0){\includegraphics[height=0.3\textheight,width=0.45\textwidth]{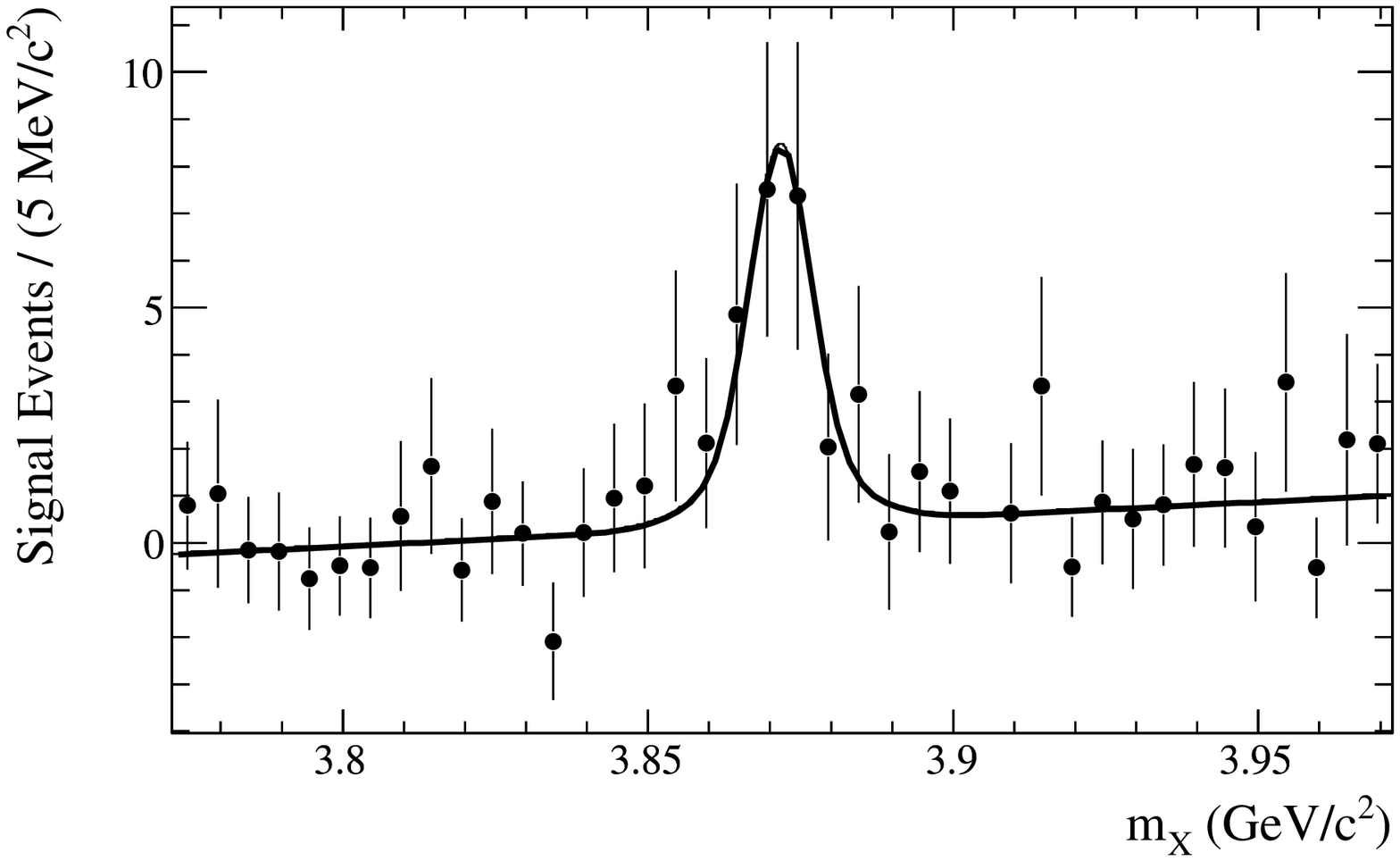}}
      \put(7.,0){\includegraphics[height=0.3\textheight,width=0.45\textwidth]{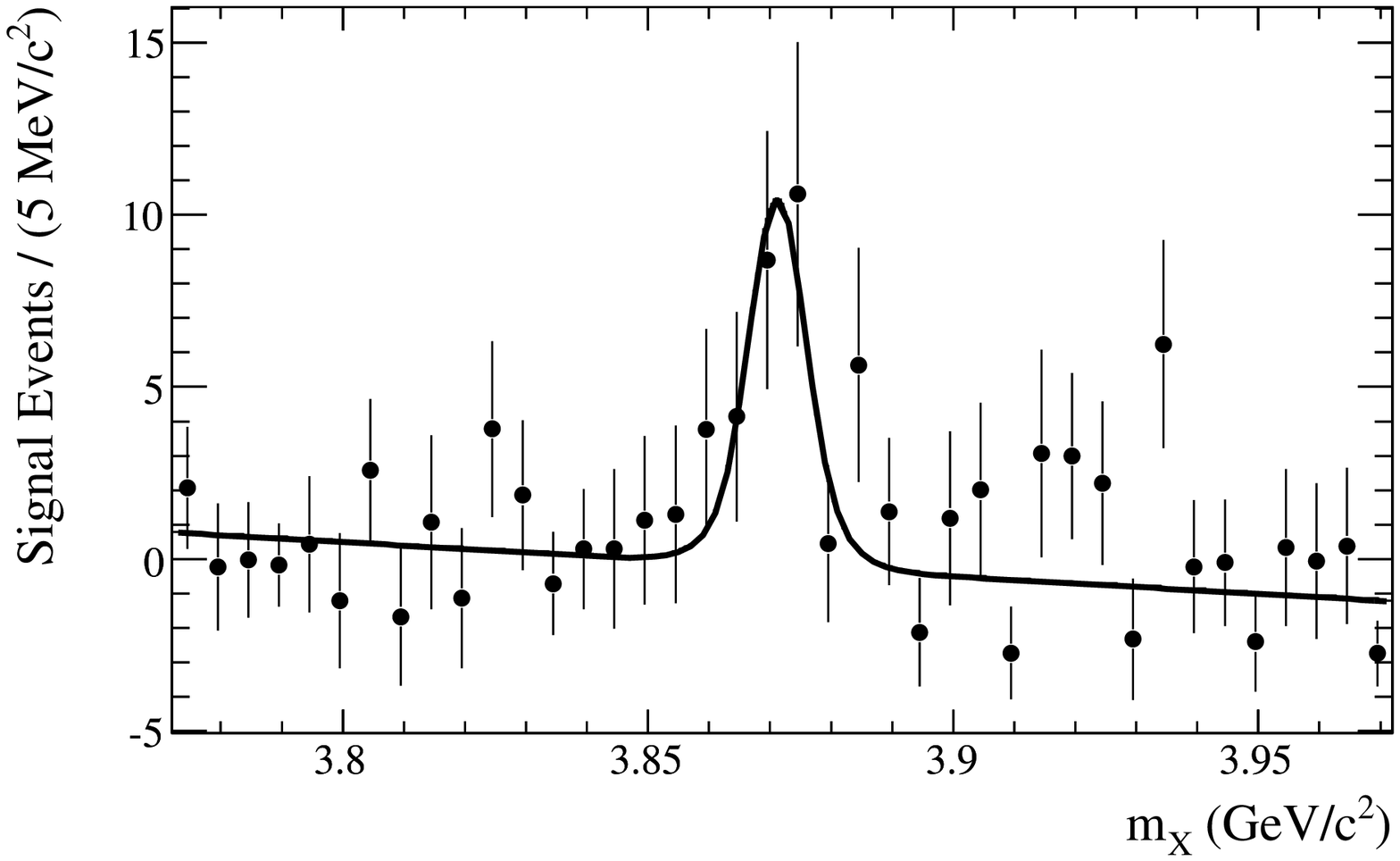}}
      \put(-0.5,5.){\bfseries \babar}
      \put(-0.5,4.7){\scriptsize \bfseries {\sl preliminary}} 
      \put(8.5,5.){\bfseries \babar}
      \put(8.5,4.7){\scriptsize \bfseries {\sl preliminary}}   
    \end{picture}
    \caption{Left (Right): The $J/\psi\gamma$ ($\psi(2S)\gamma$) mass
      distribution for $B\rightarrow X(3872)K$, $X(3872)\rightarrow
      J/\psi\gamma$ ($\psi(2S)\gamma$). The points represent the data
      and the solid curves show the results of a fit to a Breit-Wigner
      enhancement above a linear background.}
    \label{fig:x3872}
  \end{center}
\end{figure*}

The $X(3872)$, discovered by BELLE~\cite{Choi:2003ue}, was the first
of the new charmonium-like states, and has since been confirmed by
other
experiments~\cite{Acosta:2003zx,Abazov:2004kp,Aubert:2004ns}. The
$X(3872)$ was discovered in the decay mode $B\rightarrow X(3872)K$,
$X(3872)\rightarrow J/\psi\pi^+\pi^-$.  Subsequently other decay modes
have been reported such as $X(3872)\rightarrow D^0\bar{D}^{\ast
0}$~\cite{Gokhroo:2006bt,Aubert:2007rva}, and $X(3872)\rightarrow
J/\psi\gamma$~\cite{Abe:2005ix,Aubert:2006aj}. The \babar\/
Collaboration has updated the measurement of $X(3872)\rightarrow
J/\psi\gamma$ and reported a new decay mode, $X(3872)\rightarrow
\psi(2S)\gamma$, using the full \babar\ dataset of 424
fb$^{-1}$~\cite{Aubert:2008rn}. Figure~\ref{fig:x3872} shows the
corresponding evidence $J/\psi\gamma$ (left) and $\psi(2S)\gamma$
(right) mass distributions. We report branching fractions
${\cal{B}}(B^{\pm}\rightarrow X(3872)K^{\pm})\cdot
{\cal{B}}(X(3872)\rightarrow J/\psi\gamma)=(2.8\pm 0.8(stat)\pm
0.2(syst))\times 10^{-6}$ and ${\cal{B}}(B^{\pm}\rightarrow
X(3872)K^{\pm})\cdot {\cal{B}}(X(3872)\rightarrow
\psi(2S)\gamma)=(9.9\pm 2.9(stat)\pm 0.6(syst))\times 10^{-6}$. The
latter branching fraction contradicts theoretical expectation, since
it was expected to be smaller than the former~\cite{Swanson:2006st}.

\section{THE \mbox{\boldmath{$Y(3940)\rightarrow J/\psi\omega$}}}
The BELLE Collaboration reported evidence for the $Y(3940)$ in the
decay $B\rightarrow Y(3940)K$, $Y(3940)\rightarrow
J/\psi\omega$~\cite{Abe:2004zs}, with mass and the width $3943\pm
11(stat)\pm 13(syst)$ \mevcc\ and $87\pm 22(stat)\pm 26(syst)$ \mev\/,
respectively. The \babar\ Collaboration has
confirmed~\cite{Aubert:2007vj} the existence of the $Y(3940)$ using a
data sample of 348 fb$^{-1}$, but measured a lower mass
($3914.6^{+3.8}_{-3.4}(stat)\pm 2(syst)$ \mevcc\/) and smaller width
($34^{+12}_{-8}(stat)\pm 5(syst)$ \mev\/) than in the BELLE
analysis. The \babar\ ratio of $B^0$ and $B^+$ branching fractions
events for the $Y(3940)$ is $0.27^{+0.28}_ {-0.23}(stat)^{+0.04}
_{-0.01}(syst)$. The central value of this ratio is three standard
deviations below the isospin expectations, but agrees well with the
corresponding $X(3872)$ ratio from \babar\/~\cite{Aubert:2008gu}.

\section{SEARCH FOR \mbox{\boldmath{$Z(4430)^-\rightarrow J/\psi\pi^-$}} AND \mbox{\boldmath{$Z(4430)^-\rightarrow \psi(2S)\pi^-$}}}
The BELLE Collaboration reported~\cite{:2007wga} a new charmonium-like
structure, the \z\/, in the $\psi(2S)\pi^-$ system produced in the
decays $B^{-,0}\rightarrow\psi(2S)\pi^-K^{0,-}$ has generated a great
deal of interest. If confirmed, such a state must have minimum quark
content ($c\bar{c}d\bar{u}$) so that it would represent the unequivocal
manifestation of a four-quark meson state.

\begin{figure*}[!htbp]
  \begin{center}
    \setlength{\unitlength}{1.0cm}
    \begin{picture}(17,7.5)
      \put(0,0){\includegraphics[height=0.3\textheight,width=0.9\textwidth]{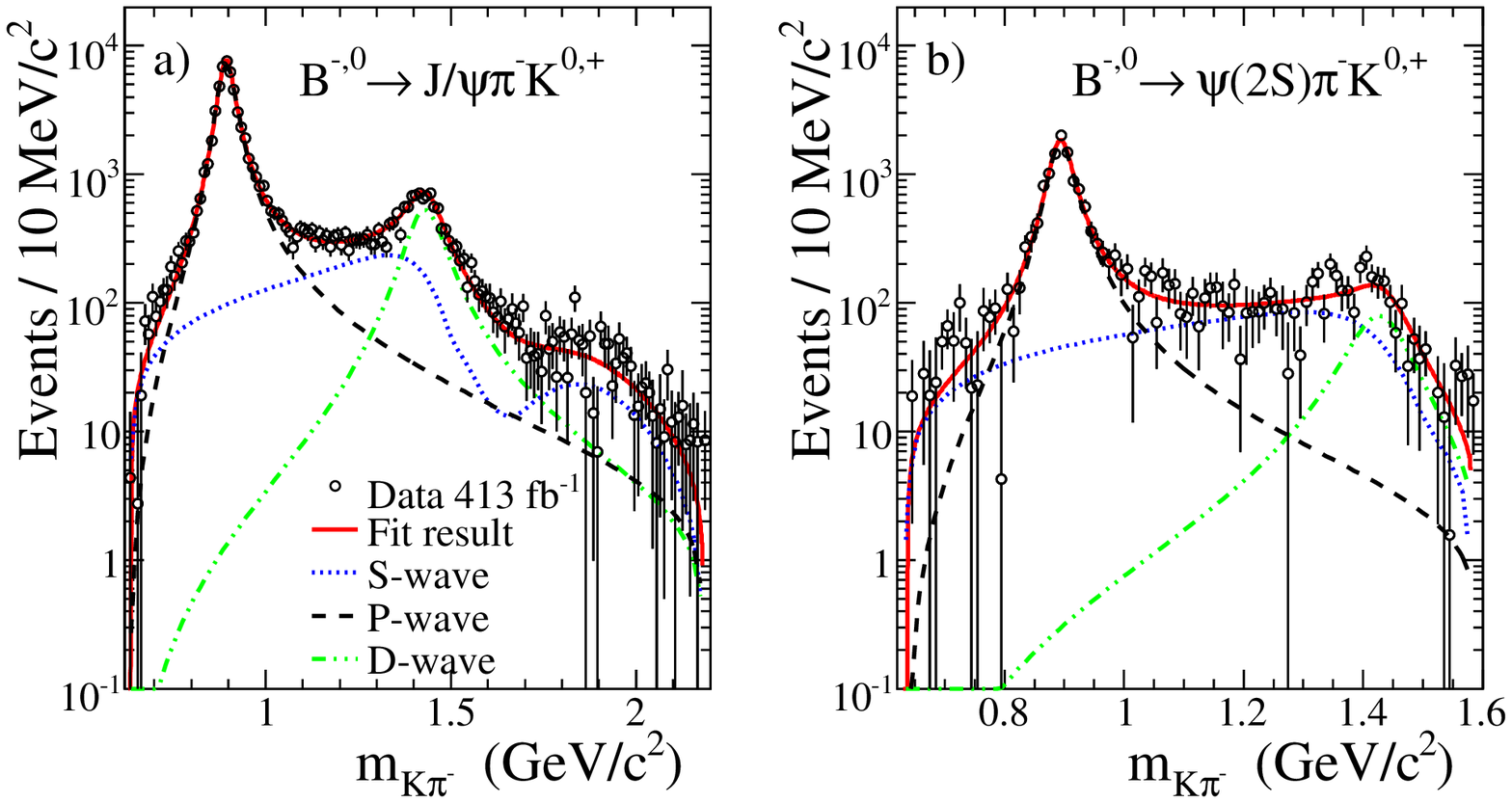}}
      \put(16.,5.5){\bfseries \babar}
      \put(16.,5.2){\scriptsize \bfseries {\sl preliminary}}                       
    \end{picture}
    \caption{The results of the fits to the \kpi\ mass distributions
      for the combined \kpi\ charge configurations (a)
      $B^{-,0}\rightarrow J/\psi\pi^- K^{0,+}$ and (b)
      $B^{-,0}\rightarrow \psi(2S)\pi^- K^{0,+}$. The individual
      intensity contributions due to $S$-, $P$-, and $D$-waves are as
      indicated.}
    \label{fig:kpi_combined}
  \end{center}
\end{figure*}

The \babar\/ Collaboration has analyzed the entire data sample
collected at the $Y(4S)$ resonance (413 fb$^{-1}$) to search for the
\z\ state in four decay modes $B\rightarrow \psi\pi^- K$, where
$\psi=J/\psi$ or $\psi(2S)$ and $K=K^0_S$ or $K^+$. The $\psi\pi^-$
mass resolution is $\sim 7 (4)$ \mevcc\ at the \z\ mass for the
$J/\psi$ ($\psi(2S)$) modes.  The $K\pi^-$ mass distributions, after
efficiency-correction and background-subtraction, for the decay modes
$B^{-,0}\rightarrow J/\psi\pi^- K^{0,-}$ and $B^{-,0}\rightarrow
\psi(2S)\pi^- K^{0,-}$ are shown in Figs.~\ref{fig:kpi_combined}(a)
and (b). The data are well-described as a sum of \kpi\ $S$-, $P$-, and
$D$-wave intensity contributions with clear enhancement of the \Ksone\
and \Kstwo\/. The \Ksone\/ mass and width values obtained are in good
agreement with the PDG values~\cite{Amsler:2008zz}.

\begin{figure*}[t]
\begin{center}
 \setlength{\unitlength}{1.0cm}
    \begin{picture}(17,7.)
      \put(0,0){\includegraphics[height=0.3\textheight,width=0.9\textwidth]{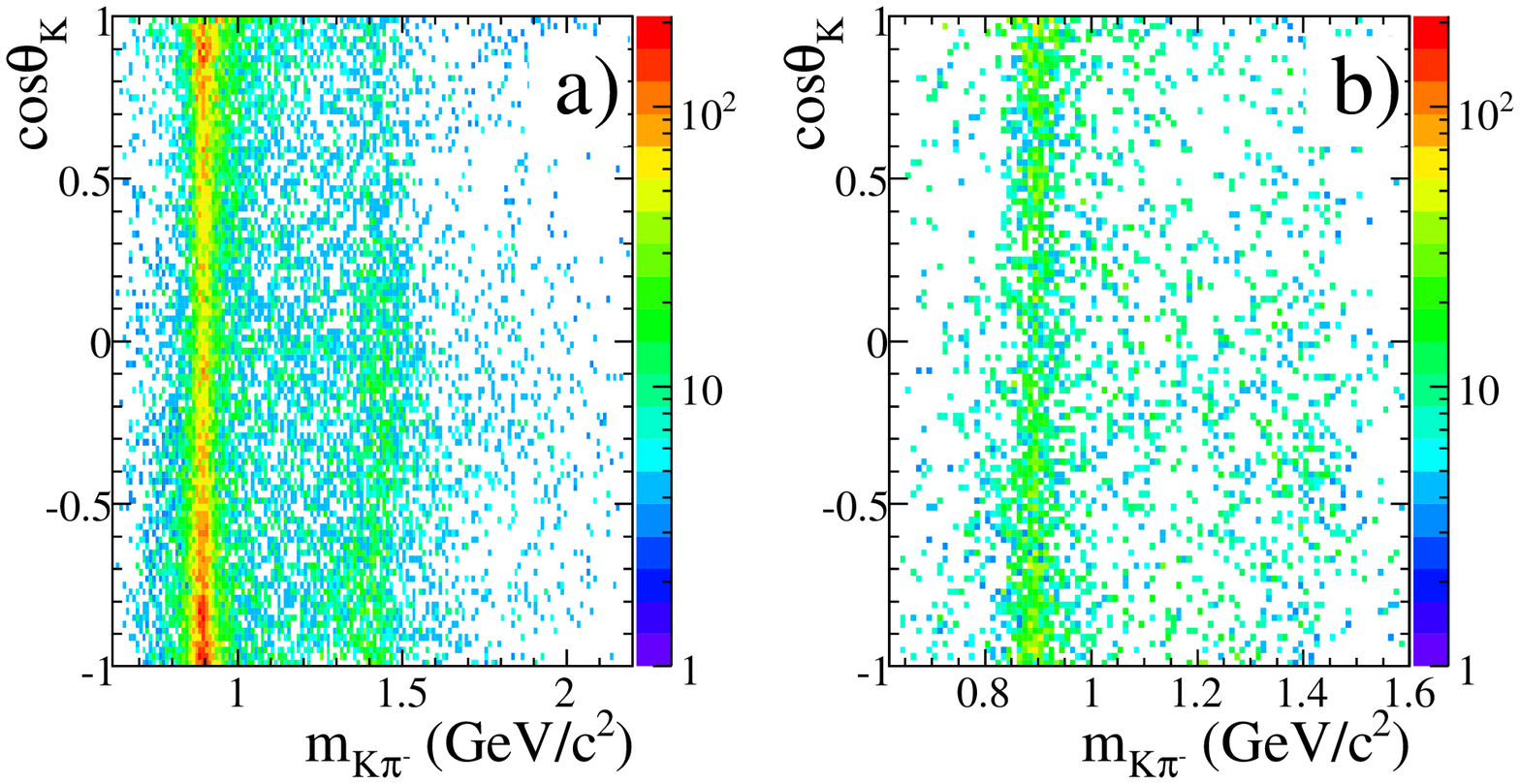}}
      \put(16.,4.5){\bfseries \babar}
      \put(16.,4.2){\scriptsize \bfseries {\sl preliminary}}                 
    \end{picture}
    \caption{The \costhk\ versus \mkpi\ for the combined decay modes
      (a) $B^{-,0}\rightarrow J/\psi\pi^- K^{0,+}$, (b)
      $B^{-,0}\rightarrow \psi(2S)\pi^- K^{0,+}$; the data samples are
      corrected for efficiency.}
    \label{fig:dp3}
\end{center}
\end{figure*}

\begin{figure*}[t]
  \begin{center}
    \setlength{\unitlength}{1.0cm}
    \begin{picture}(17,9.)
      \put(0,0){\includegraphics[height=0.4\textheight,width=0.9\textwidth]{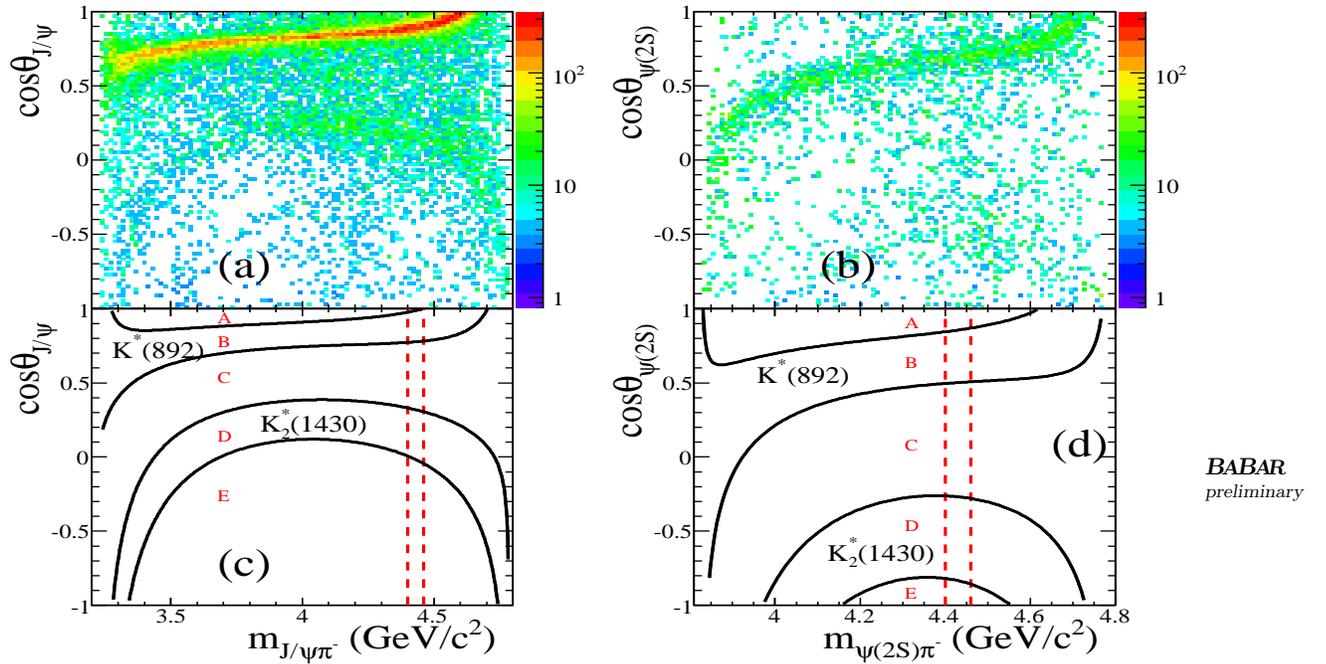}}
      \put(16.,2.5){\bfseries \babar}
      \put(16.,2.2){\scriptsize \bfseries {\sl preliminary}}           
    \end{picture}
    \caption{The $\cos\theta_{\psi}$ versus $m_{\psi\pi^-}$
      rectangular Dalitz plots for (a) $B^{-,0}\rightarrow J/\psi\pi^-
      K^{0,+}$, and (b) $B^{-,0}\rightarrow \psi(2S)\pi^- K^{0,+}$;
      (c) and (d), the corresponding plots indicating the loci of the
      \Ksone\ and \Kstwo\ resonance bands defined in the text; regions
      A-E, as defined in the text, are indicated. The dashed vertical
      lines indicate the mass range $4.400<m_{\psi\pi^-}<4.460$ \gevcc\/.}
    \label{fig:dp1}
  \end{center}
\end{figure*}

We represent the $K\pi^-$ mass dependence of the angular structure in
the $K\pi^-$ angular distribution at a given $m_{K\pi^-}$ in terms of
Legendre polynomials, $P_l(\cos\theta_K)$, where the angle $\theta_K$
is between the $K$ in the \kpi\ rest frame and the \kpi\ direction in
the $B$ rest frame. The $\cos\theta_K$ versus \mkpi\ distributions for
the $J/\psi$ and $\psi(2S)$ samples are shown in Fig.~\ref{fig:dp3},
and both exhibit backward-forward asymmetry in $\cos\theta_K$. The
\kpi\ mass and angular structure has a significant impact on the
corresponding \psipi\ mass distribution. This structure is shown in
Fig.~\ref{fig:dp1}, where we plot $\cos\theta_\psi$ versus
$m_{\psi\pi^-}$; $\theta_\psi$ is the angle between the $\psi$ in the
$\psi\pi^-$ rest frame, and the \psipi\ direction in the $B$ rest
frame. The \Ksone\ bands are observed and a \Kstwo\ band can be seen
in Fig.~\ref{fig:dp1}(a). The regions corresponding to the the \kpi\
mass intervals (A) below the \Ksone\/, (B) within 100 \mevcc\ of the
\Ksone\ nominal mass, (C) between the \Ksone\ and \Kstwo\/, (D) within
100 \mevcc\ of the \Kstwo\ nominal mass, and (E) above the \Kstwo\ are
labeled.

The \z\ signal in Ref.~\cite{:2007wga} corresponds to regions A, C, and
E combined; Fig.~\ref{fig:dp1}(d) shows that, at the \z\ mass
position, only about half of the decay angular distribution is then
being selected.

\begin{figure*}[t]
\begin{center}
  \setlength{\unitlength}{1.0cm}
  \begin{picture}(17,12.0)
    \put(0,0){\includegraphics[height=0.5\textheight,width=0.9\textwidth]{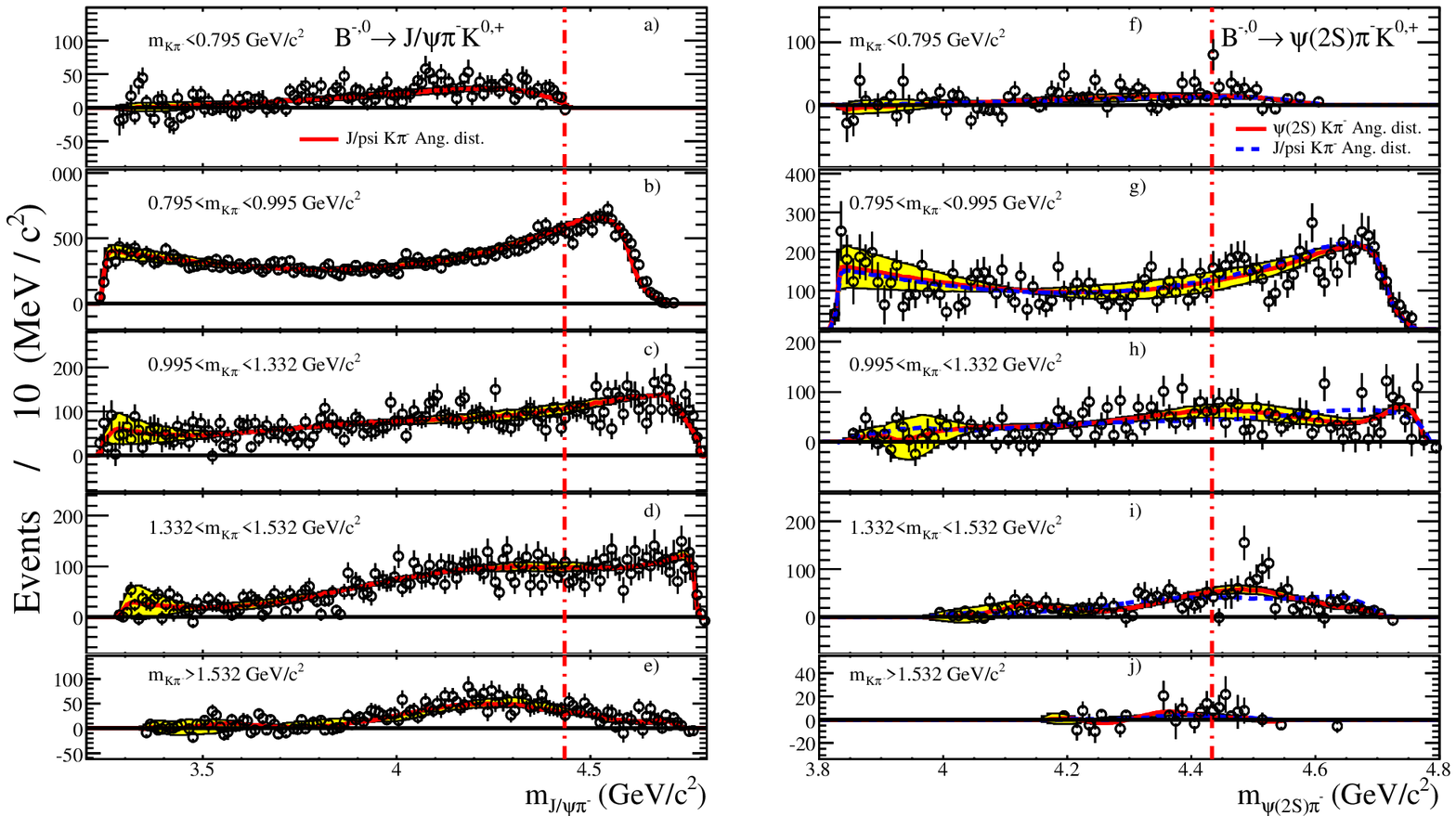}}
    \put(16.0,6.5){\bfseries \babar}
    \put(16.0,6.2){\scriptsize \bfseries {\sl preliminary}}     
  \end{picture}
\caption{The \psipi\ mass distributions in regions A-E of \kpi\ mass
for the combined decay modes (a-e) $B^{-,0}\rightarrow J/\psi\pi^-
K^{0,+}$, and (f-j) $B^{-,0}\rightarrow \psi(2S)\pi^- K^{0,+}$; the
solid curves and shaded bands are described in the text. In (f-j), the
dot-dashed curves are obtained using \kpi\ normalized moments for
$B^{-,0}\rightarrow J/\psi\pi^- K^{0,+}$, instead of those from
$B^{-,0}\rightarrow \psi(2S)\pi^- K^{0,+}$; the dashed vertical lines
indicate $m_{\psi\pi^-}=4.433$ \gevcc\/.}
\label{fig:ranges}
\end{center}
\end{figure*}

The \psipi\ mass distributions in the five \kpi\ mass ranges (A-E) are
shown in Fig.~\ref{fig:ranges} for the samples with $J/\psi$ and
$\psi(2S)$ candidates. The solid curves represent the \kpi\ reflection
into the \psipi\ mass distribution, taking into account the \kpi\ mass
structure through the fit functions of Fig.~\ref{fig:kpi_combined},
and the $\cos\theta_K$ dependence via the normalized Legendre
polynomial moments. The bands indicate the uncertainties resulting
from the uncertainties in the Legendre moments. No significant
enhancement of the data and the solid curves is observed at the \z\
mass in any \kpi\ region.

\begin{figure*}[t]
  \begin{center}
    \setlength{\unitlength}{1.0cm}
    \begin{picture}(17,12)
    \put(0,0){\includegraphics[height=0.5\textheight,width=0.9\textwidth]{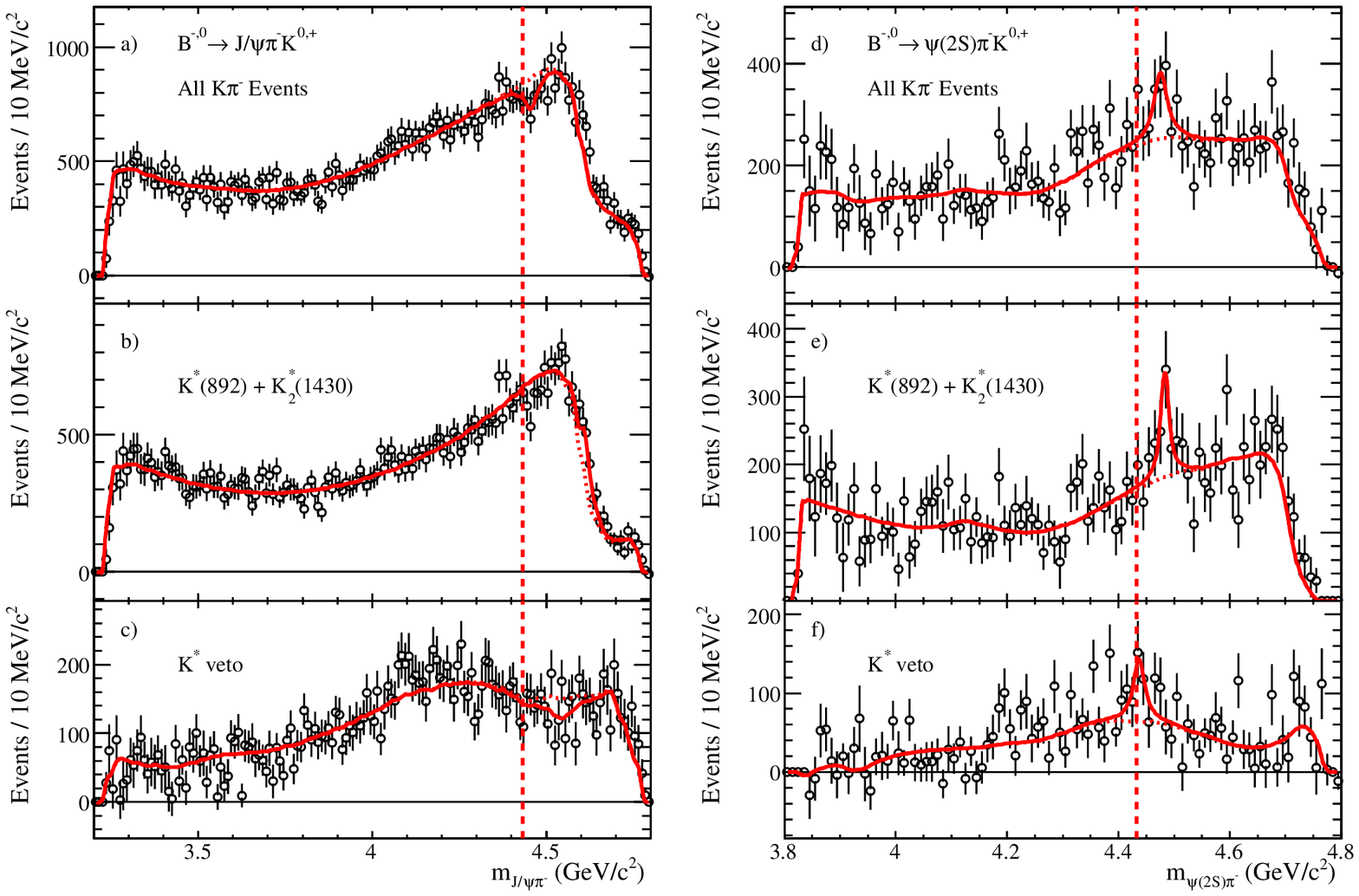}}
    \put(16.0,6.5){\bfseries \babar}
    \put(16.0,6.2){\scriptsize \bfseries {\sl preliminary}} 
    \end{picture}
\caption{The results of the fits to the corrected mass distributions,
(a)-(c) for \jpsipi\/, and (d)-(f) for \psitwospi\/. The curves are
described in the text; the dashed vertical lines indicate
$m_{\psi\pi^-}=4.433$ \gevcc\/.}
\label{fig:babar_cor}
\end{center}
\end{figure*}

In Fig.~\ref{fig:babar_cor} we show fits to the \psipi\ mass
distributions in which the \kpi\ background shape is fixed and an
$S$-wave Breit Wigner(BW) is used as signal function. The solid curves
represent the fit results while the dashed curves show the \kpi\
background. The BW parameters are free in the
fits. Figures~\ref{fig:babar_cor}(a) and (d) are for the entire-data
samples; Figures~\ref{fig:babar_cor}(b) and (e) are for the $K^{\ast}$
regions (B plus D), and Figs~\ref{fig:babar_cor}(c) and (f) are for
regions A, C, and E combined (the BELLE selection). For \jpsi\
samples, no evidence for any enhancements is obtained. For the
\psitwos\ data small signals are obtained, but their significance is
only in the $2-3\sigma$ range, and in Fig.~\ref{fig:babar_cor}(d) and
(e) the fitted mass is significantly different from the BELLE
value. In Fig.~\ref{fig:babar_cor}(f), the signal mass and width are
consistent with the BELLE values, but the signal significance is only
$1.9\sigma$. We conclude that the \babar\ data provide no significant
evidence for the existence of the \z\/.


\begin{thebibliography}{9}   
\bibitem{Aubert:2005rm} B.~Aubert {\it et al.},  [\babar\ Collaboration],  Phys.\ Rev.\ Lett.\  {\bf 95}, 142001 (2005).
\bibitem{:2008ic} B.~Aubert {\it et al.}, [\babar\ Collaboration], arXiv:0808.1543 [hep-ex].
\bibitem{:2007sj} C.~Z.~Yuan {\it et al.},  [BELLE Collaboration],  Phys.\ Rev.\ Lett.\  {\bf 99}, 182004 (2007).
\bibitem{Choi:2003ue} S.~K.~Choi {\it et al.},  [BELLE Collaboration],  Phys.\ Rev.\ Lett.\  {\bf 91}, 262001 (2003).

\bibitem{Acosta:2003zx} D.~E.~Acosta {\it et al.},  [CDF II Collaboration],Phys.\ Rev.\ Lett.\  {\bf 93}, 072001 (2004).
\bibitem{Abazov:2004kp} V.~M.~Abazov {\it et al.},  [D0 Collaboration],  Phys.\ Rev.\ Lett.\  {\bf 93}, 162002 (2004).
\bibitem{Aubert:2004ns} B.~Aubert {\it et al.},  [\babar\ Collaboration],  Phys.\ Rev.\  D {\bf 71}, 071103 (2005)

\bibitem{Gokhroo:2006bt} G.~Gokhroo {\it et al.}, [BELLE Collaboration], Phys.\ Rev.\ Lett.\  {\bf 97}, 162002 (2006).
\bibitem{Aubert:2007rva} B.~Aubert {\it et al.},  [\babar\ Collaboration], Phys.\ Rev.\  D {\bf 77}, 011102 (2008).
\bibitem{Abe:2005ix} K.~Abe {\it et al.}, [BELLE Collaboration], arXiv:0505037 [hep-ex].
\bibitem{Aubert:2006aj} B.~Aubert {\it et al.}, [\babar\ Collaboration],  Phys.\ Rev.\  D {\bf 74}, 071101 (2006).
\bibitem{Aubert:2008rn} B.~Aubert {\it et al.},  [The \babar\ Collaboration],  arXiv:0809.0042 [hep-ex].
\bibitem{Swanson:2006st} E.~S.~Swanson,  Phys.\ Rept.\  {\bf 429}, 243 (2006).
\bibitem{Abe:2004zs} K.~Abe {\it et al.},  [BELLE Collaboration],  Phys.\ Rev.\ Lett.\  {\bf 94}, 182002 (2005).
\bibitem{Aubert:2007vj} B.~Aubert {\it et al.},  [\babar\ Collaboration],  Phys.\ Rev.\ Lett.\  {\bf 101}, 082001 (2008).
\bibitem{Aubert:2008gu} B.~Aubert {\it et al.},  [\babar\ Collaboration],  Phys.\ Rev.\  D {\bf 77}, 111101 (2008).
\bibitem{:2007wga} S.-K.~Choi {\it et al.},  [BELLE Collaboration],  Phys.\ Rev.\ Lett.\  {\bf 100}, 142001 (2008).
\bibitem{Amsler:2008zz} C.~Amsler {\it et al.},  [Particle Data Group],  Phys.\ Lett.\  B {\bf 667} (2008) 1.

\end{thebibliography}
\end{document}